\documentstyle[12pt]{article}
\begin{document}
\baselineskip=20pt
\begin{center}
{\large\bf Dynamical theory of quantum chaos or\\
a hidden random matrix ensemble?}
\end{center}

\vspace{1cm}

In a recent Letter \cite{1} Altland and Zirnbauer claim that they rigorously
proved the {\it complete} analogy between a (classically chaotic) dynamical
system and disordered (random) solids.  The main purpose of this Comment is
to show that,in fact, the theory \cite{1} does not represent at least some
characteristic dynamical features of the quantum kicked rotator (QKR), an
example chosen in \cite{1} for comparison with the theory.

The phenomenon of quantum suppression of classically chaotic diffusion was
pointed out long ago \cite{2}. Later on, a large number of papers appeared
in which this phenomenon was numerically, analytically and even
experimentally \cite{3} confirmed. In particular, an important step was the
correspondence found between the QKR and the 1D disordered model \cite{4,5}.
However, in spite of this and in spite of all mathematical efforts , so far,
a rigorous proof of quantum localization in the QKR is still lacking. The
main mathematical difficulty lies in the fact that, for a rigorous proof of
localization, the theoretic number properties of the period $\tau$ are
crucial. As a matter of fact, rigorous statements can be made only in the
opposite direction. Indeed, it was analytically shown \cite{6} that for
rational values of the parameter $T=\tau /4 \pi$ no localization takes place
(the so-called "quantum resonance").  Moreover, it was rigorously proven
\cite{7} that localization is absent also for a non-empty set of irrational
values of $T$.  It is not known how `large' is this set (even if we may
guess that its Lebegue measure is zero) and if and how it depends on the
perturbation strength $k$.  Even for a typical (with probability 1)
irrational $T$, when the whole everywhere dense set of quantum resonances is
still insufficient to destroy localization, the eigenfunctions may be,
nevertheless, essentially modified if the detuning from resonance is small
enough. The measure of such $T$ is typically small but finite \cite{6}. As
far as we understand, the paper \cite{1} does not take into account this
problem which constitutes an important feature and difficulty of the QKR.

In our opinion, the important dynamical features mentioned above were lost
in the approximate calculations following the exact functional (6) in ref.
[1].  The following simple argument can demonstrate our main point.  By
direct averaging over the parameter $\omega_0$ (we use the same notations as
in \cite{1}) one easily finds
$$Q_{1,2;2,1}(\omega)\equiv
\langle\;\langle l_1|G^+(\omega_+)|l_2\rangle\,
\langle l_2|G^-(\omega_-)|l_1\rangle\;\rangle_{\omega_0} = $$
\begin{equation}\label{1}
\sum_{n=0}^{\infty}\exp[in(\omega\tau+i0)]\,
|\langle l_1|U^n|l_2 \rangle|^2\;.
\end{equation}
This \underline{exact} expression contains matrix elements of the dynamical
Floquet operator $U$ so that all dynamical peculiarities of QKR survive the
$\omega_0$-averaging. We reckon that eq. (6) in ref. [1] gives an integral
representation of the  quantity (\ref{1}) and therefore carries the same
full dynamical information. Let us now consider, for example, the simplest
quantum resonance with $\tau=4\pi$. In this case $\langle l_1|U^n|l_2 \rangle=
\langle l_1|\exp[-i(nk\,cos\theta)]|l_2 \rangle$
and (\ref{1}) reduces to
\begin{equation}\label{3}
Q_{1,2;2,1}(\omega) = \sum_{n=0}^{\infty}\exp[in(\omega\tau+i0)]\;
J^2_{|l_1-l_2|}(nk)
\end{equation}
where $J_l$ stands for the Bessel function.  Using cosine asymptotics of the
Bessel function one readily estimates the singular part of (\ref{3}) at the
infrared limit $\omega\rightarrow 0$ as
\begin{equation}\label{4}
Q_{1,2;2,1}(\omega) = \frac{2}{\pi k}\;\ln\frac{i}{\omega\tau+i0}
\end{equation}
Contrary to the localization regime, the r.h.s in (\ref{4}) does not depend
on the distance $|l_1-l_2|$. Such resonance regimes of motion (as well as
the nearly resonant values of T) are not reproduced by the calculation
reported in \cite{1}. They ``mysteriously" disappear from the theory between
eqs. (6) and (8). This leads us  to suspect that some hidden additional
statistical assumption has been made in \cite{1} which remains uncleared.
Most likely, the very structures of the principal configurations in (6) must
be quite different depending on how close to rationals is parameter T. This
delicate problem is completely ignored in \cite{1}.

An additional remark concerns the dependence on the "symmetry breaking"
parameter $a$. It is trivially seen that the introduction of this
parameter results only in the phase transformation
\begin{equation}\label{a}
\langle l_1|U^n|l_2 \rangle \rightarrow
\exp[i(l_1-l_2)\,a]\;\langle l_1|U^n|l_2 \rangle
\end{equation}
of the matrix elements of an arbitrary power $n$ of the Floquet operator.
Therefore, this parameter completely disappears from eq. (\ref{1}) and by no
means can influence the localization length. The presence of this parameter
in the final expression (9) in Ref. \cite{1} is fully due to the angular
discretization used. An analysis of the dependence of the
localization length on symmetry breaking parameters in the generalized
QKR is given in \cite{8}.

Finally, it is probably worth just mentioning the kicked Harper model which
exhibits the same classical diffusion as the kicked rotator but which has an
extremely rich quantum behaviour; in particular, delocalization can take
place for any value, rational or irrational, of the kick period T. As
discussed in \cite{9} this model clearly shows how subtle is the problem to
understand the features of classical chaotic dynamics which lead to quantum
diffusion or localization. This problem remains open.

In conclusion, the theory \cite{1}, in its present form
may be expected to be valid when standard band
random matrix approach can be ad hoc used. 
However, it fails to take into account dynamical features
which go beyond the random matrix theory description.

We are indebted to B.Chirikov for an important impact by his numerous and
deep remarks. We appreciate also interesting discussions with Ya.V. Fyodorov
A.D. Mirlin and O.V.Zhirov.\\

PACS numbers 03.65.Bz, 05.40.+j, 05.45.+b


\vspace{0.5cm}

Giulio Casati

\hspace{1cm} International Center for the Study of Dynamical Systems

\hspace{1cm} University of Milano at Como

\hspace{1cm} Via Castelnuovo, 7 - 22100 COMO Italy



\vspace{5mm}

F.M.Izrailev and V.V.Sokolov

\hspace{1cm} Budker Institute of Nuclear Physics

\hspace{1cm} 630090 Novosibirsk, Russia

\vspace{5mm}


\end{document}